%% file: paper.tex
\documentclass[aps,prb,twocolumn,a4paper,10pt,superscriptaddress]{revtex4-1}

\input{preamble}

\newcommand{\figL}[1]{\vcenter{\hbox{\includegraphics[scale=0.4]{./#1.pdf}}}}
\newcommand{\paragraphL}[1]{\paragraph*{{#1}.}\addcontentsline{toc}{section}{#1}}

\date{\today}

\begin{document}

\title{Excitations and the tangent space of projected entangled-pair states}

\author{Laurens Vanderstraeten}
\author{Micha\"el Mari\"en}
\affiliation{Ghent University, Department of Physics and Astronomy, Krijgslaan 281-S9, B-9000 Gent, Belgium}
\author{Frank Verstraete}
\affiliation{Ghent University, Department of Physics and Astronomy, Krijgslaan 281-S9, B-9000 Gent, Belgium}
\affiliation{Vienna Center for Quantum Science, Universit\"at Wien, Boltzmanngasse 5, A-1090 Wien, Austria}
\author{Jutho Haegeman}
\affiliation{Ghent University, Department of Physics and Astronomy, Krijgslaan 281-S9, B-9000 Gent, Belgium}

\begin{abstract}
We introduce tangent space methods for projected entangled-pair states (PEPS) that provide direct access to the low-energy sector of strongly-correlated two-dimensional quantum systems. More specifically, we construct a variational ansatz for elementary excitations on top of PEPS ground states that allows for computing gaps, dispersion relations and spectral weights directly in the thermodynamic limit. Solving the corresponding variational problem requires the evaluation of momentum transformed two-point and three-point correlation functions on a PEPS background, which we can compute efficiently by introducing a new contraction scheme. As an application we study the spectral properties of the magnons of the Affleck-Kennedy-Lieb-Tasaki model on the square lattice and the anyonic excitatons in a perturbed version of Kitaev's toric code.
\end{abstract}

\maketitle

In the last decades, low-dimensional quantum systems have been at the forefront of both theoretical and experimental physics. With the reduced dimensionality allowing for stronger quantum correlations, these systems show an extreme variety of exotic phenomena but are notoriously difficult to simulate \cite{Fradkin2013}. In particular, the low-energy physics of such systems typically cannot be apprehended by perturbing around some free or otherwise exactly solvable point. 
\par For one-dimensional quantum spin chains, the advent of the density matrix renormalization group (DMRG) \cite{White1992} has proven revolutionary. DMRG has developed into the de facto standard method for reliably and efficiently probing the low-energy behaviour of strongly correlated systems in one dimension. A better understanding of this success was realized through the identification of the variational class over which DMRG optimizes as \emph{matrix product states} (MPS) \cite{Fannes1989, *Fannes1992, Ostlund1995a, *Rommer1997a}. More specifically, it was understood how the entanglement structure of MPS allows for a natural parametrization of the low-energy states of gapped, local one-dimensional quantum systems \cite{Hastings2007, *Verstraete2006}.
\par The characterizing feature of the entanglement structure in gapped local quantum systems is the observed area law for entanglement entropy \cite{Eisert2010a}. With this insight the natural extension of MPS to higher dimensions is given by the class of \emph{projected entangled-pair states} (PEPS) \cite{Verstraete2004b, *Verstraete2006a, Verstraete2008a, *Orus2013}. This set of states has shown to capture the low-energy physics of several interesting systems in two dimensions and is competing with more established methods in determining their ground-state properties \cite{Jordan2008, Murg2007, *Dusuel2011, *Poilblanc2014, *Corboz2014, *Corboz2014a, *Corboz2015}. Yet, in contrast to the one-dimensional case, the PEPS simulation of two-dimensional systems is computationally challenging, thus making the development of new algorithms highly desirable. Furthermore, existing PEPS algorithms focus exclusively on capturing ground state wave functions, whereas experimentally relevant low-energy properties such as excitation spectra remain inaccessible.
\par In this paper we initiate a new set of methods based on the \emph{tangent space} of the PEPS manifold of states, a concept which has proven extremely versatile in the context of matrix product states \cite{Haegeman2013b}. We apply this methodology for constructing a variational ansatz for elementary excitations on top of PEPS ground states. Combined with the versatility of the PEPS as ground state ansatz, our method provides a flexible and mostly unbiased approach (aside from entanglement considerations in the PEPS ansatz) towards extracting the low-energy spectrum of a strongly correlated two-dimensional quantum system. We explain the main ingredients for successfully applying the variational principle, for which we need to introduce a new PEPS contraction scheme. As an application we study the excitation spectrum of the two-dimensional Affleck-Kennedy-Lieb-Tasaki (AKLT) model on a square lattice, and of a perturbed version of the toric code model, where we show the ability to access single topological excitations (anyons) directly.

\paragraphL{Elementary excitations} We start from one of the central insights of condensed-matter physics that the low-energy properties of quantum systems can be understood in terms of elementary excitations or quasi-particles \cite{Anderson1963}. Experimental observables such as dynamical correlation functions, real-time evolution or low-temperature properties can be understood from this picture of weakly-interacting particles on a non-trivial vacuum state. This insight can be materialized within the PEPS formalism by first identifying the PEPS class of states as a variational manifold, embedded in the full Hilbert space, that captures the ground state structure of local Hamiltonians. Analoguous to e.g. the manifold of Slater determinants for the Hartree-Fock approximation of a fermionic ground state \cite{Rowe1980}, or the manifold of matrix product states \cite{Haegeman2014} for ground states of quantum spin chains, the excitations on top of such a ground state are obtained as linear perturbations living in the tangent space to the manifold rather than the manifold itself. This gives rise to a set of methods known as post-Hartree-Fock \cite{Helgaker2014} or post-MPS methods \cite{Haegeman2013b, Kinder2011, *Wouters2013}, respectively. Here, we develop ``post-PEPS'' methods by constructing elementary excitations within the tangent space of PEPS ground states.

\paragraphL{PEPS and variational excitations} Consider a two-dimensional quantum lattice system in the thermodynamic limit. A translation-invariant PEPS \cite{Verstraete2004b, *Verstraete2006a} can be parametrized by a single five-legged tensor $A^s_{u,r,d,l}$ as
\begin{equation*}
\ket{\Psi(A)} = \sum_{\{s\}=1}^d \mathsf{C}_2 (\{A^s\}) \ket{\{s\}}
\end{equation*}
where $\mathsf{C}_2(\dots)$ denotes the contraction of an infinite two-dimensional network of $A$ tensors, which is more conveniently represented pictorially as
\begin{equation} \label{eq:PEPS}
\ket{\Psi(A)} = \figL{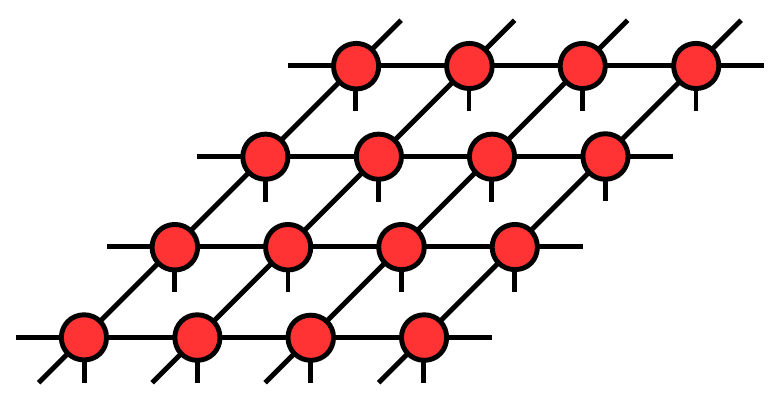}.
\end{equation}
Here, we have represented the tensor $A$ as
\begin{equation*}
A^s_{u,l,d,r} = \figL{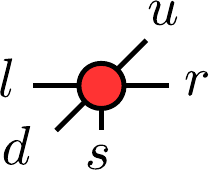}.
\end{equation*}
The dimension of the four \emph{virtual} legs ($u,r,l,d$) is called the bond dimension $D$ of the PEPS and serves as a refinement parameter of the variational class, whereas the leg $s$ has the local physical dimension. In diagrams such as in Eq.~\eqref{eq:PEPS} whenever two legs are connected the two corresponding indices are contracted (i.e. identified and summed over), such that $\ket{\Psi(A)}$ is obtained by contracting virtual legs of neighbouring tensors \footnote{In this paper we restrict to the square lattice, but our framework can be readily extended to more general lattices.}.
\par Computing the norm or a local expectation value of a PEPS involves the contraction of an infinite tensor network. Indeed, denoting the ``double layer'' tensor $a$ as
\begin{equation*}
a_{uu',rr',dd',ll'} = \sum_{s} A^s_{uldr} \otimes \overline{A}^s_{u'l'd'r'} = \figL{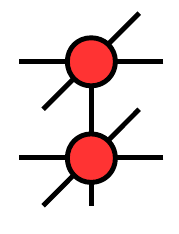} = \figL{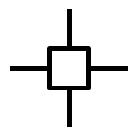},
\end{equation*}
we can represent the norm of the state in ``top view'' as
\begin{equation} \label{eq:norm}
\braket{\Psi(A)|\Psi(A)}= \figL{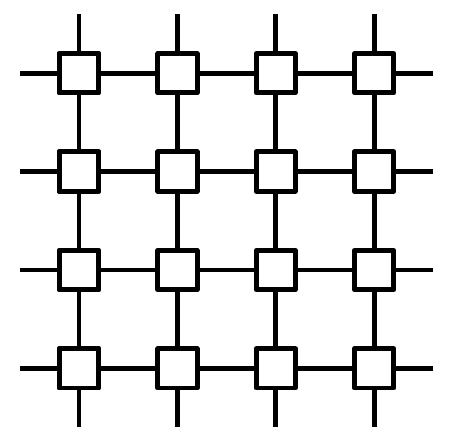}.
\end{equation}
This infinite contraction cannot be performed exactly \cite{Schuch2007, Lubasch2014a} but, by relying on MPS-style algorithms, a precise approximation can be computed efficiently, as discussed below. With a slight modification, the same algorithms can be used to compute the expectation value of local observables, such as the energy density, and therefore suffice to implement the variational principle.
\par Suppose we have found a translation-invariant PEPS representation for the ground state of a local Hamiltonian $\hat{H}$. The tangent space ansatz for elementary excitations is given by
\begin{equation} \label{eq:ansatz}
\ket{\Phi_{\kappa_x\kappa_y}(B)} = \sum_{m,n} \e^{i(\kappa_xm+\kappa_yn)} \\ \figL{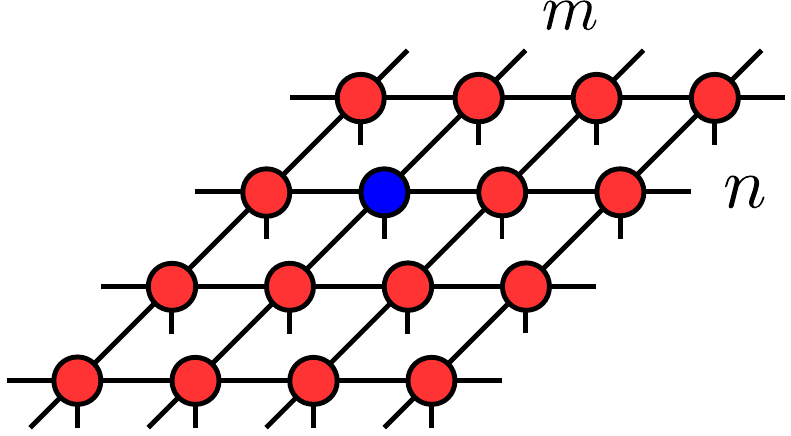}
\end{equation}
where the red tensors correspond to those of the optimal ground state and only the blue circle represents a new tensor $B$ at site $(m,n)$. This ansatz thus represents the momentum superposition of a local perturbation, which has a finite width (determined by the PEPS bond dimension) by acting via the virtual degrees of freedom. It encompasses the Feynman-Bijl ansatz \cite{Feynman1953a, *Feynman1956, *Girvin1985, *Girvin1986, Arovas1988} and its validity for gapped excitations can be rigorously motivated \cite{Haegeman2013a}. 
\par Once we have defined the variational subspace, we can minimize the energy in order to find the best approximation to the true excitation. Since the subspace is linear, this gives rise to the Rayleigh-Ritz problem
\begin{equation}\label{eq:rayleighritz}
\mathsf{H}_\text{eff}^{\kappa_x\kappa_y} \vect{B} = \omega \mathsf{N}_\text{eff}^{\kappa_x\kappa_y} \vect{B}
\end{equation}
where $\mathsf{H}_\text{eff}^{\kappa_x\kappa_y}$ and $\mathsf{N}_\text{eff}^{\kappa_x\kappa_y}$ are the effective Hamiltonian and norm matrices, which can be interpreted as carrying the one-particle dynamics on top of a PEPS background. The lowest eigenvalue $\omega$ corresponds to the lowest excitation energy at momentum $(\kappa_x,\kappa_y)$. By repeating this procedure for different momenta, we can obtain the dispersion relation for all one-particle excitations in the system. Moreover, with an expression for the wave functions, we can straightforwardly compute their spectral weights. 

\paragraphL{Effective environments} The matrix elements of the effective Hamiltonian and normalization matrices contain Fourier transformed two- and three-point functions and cannot be computed efficiently using the current PEPS contraction schemes for local expectation values. PEPS contraction schemes try to capture the effect of all the surrounding tensors in an effective environment using an approximate model. Depending on the situation, the most efficient schemes either model the environment as a matrix product state \cite{Jordan2008,Lubasch2014a} or make use of the idea of a corner transfer matrix \cite{Baxter1968, *Baxter1982, Nishino1996, *Nishino1997a, *Nishino1998, Orus2009, *Orus2012, *Corboz2010b}. For the problem at hand, we combine ideas from both schemes to construct a new  contraction method.
\par Let us first review the most straightforward algorithm for contracting the network Eq.~\eqref{eq:norm}, which consists of running through the network in one direction, sequentially applying a row of tensors, and approximating the boundary as an MPS in every step. In the infinite case, where we have an infinite number of rows, this algorithm boils down to finding a translation invariant MPS representation of the fixed point of the \emph{linear transfer matrix} $\mathcal{T}$, which can be pictorially represented as
\begin{equation*}
\mathcal{T} = \figL{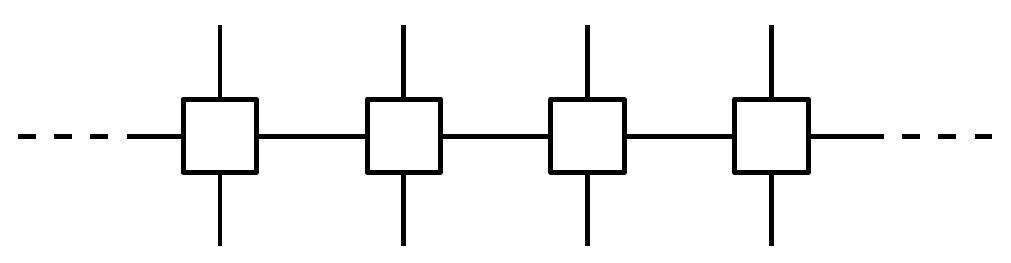}.
\end{equation*}
Once we have found this fixed point and rescaled the PEPS tensor such that the largest eigenvalue of $\mathcal{T}$ is 1 (corresponding to a normalized PEPS), the computation of local expectation values is analoguous to well-established and efficient MPS algorithms. One can also construct approximate eigenstates for the next eigenvalues of the transfer matrix $\mathcal{T}$ and infer from it the qualitative features of the physical excitation spectrum \cite{Zauner2015, Haegeman2014b}. In particular, the second largest eigenvalue $\lambda^{(2)}_\mathcal{T}$ is related to the correlation length of the PEPS as $\xi=-1/\log|\lambda^{(2)}_\mathcal{T}|$.
\par The computation of general $n$-point functions are more difficult (unless the $n$ points are aligned along one of the main axes), but are a necessary ingredient in the construction of the effective matrices in Eq.~\eqref{eq:rayleighritz}. We therefore introduce a new contraction scheme: the central idea is that, instead of contracting Eq.~\eqref{eq:norm} top-down linearly, we run through the lattice diagonally by sequentially applying a corner-shaped  transfer matrix $\mathcal{C}$ 
\begin{equation*}
\mathcal{C} = \figL{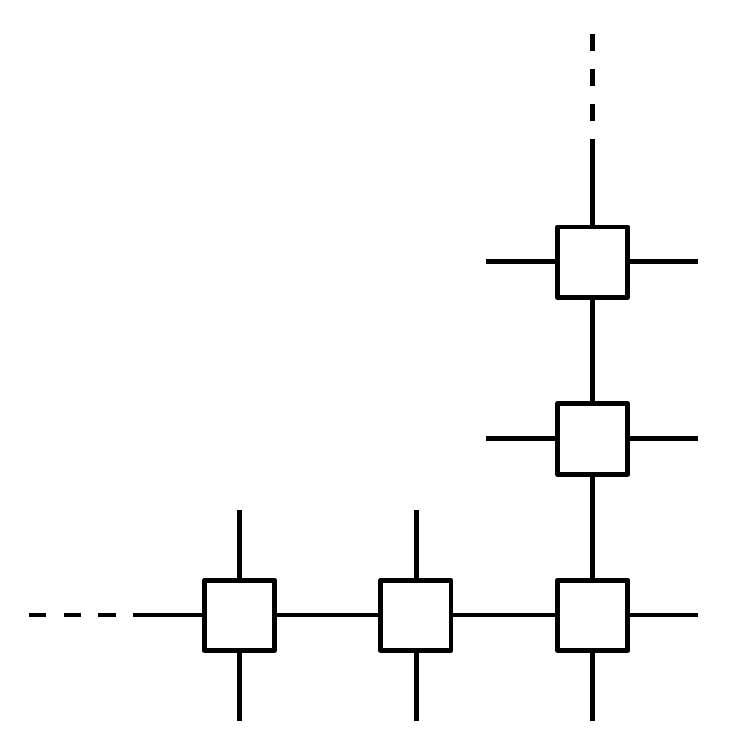}.
\end{equation*}
We can again approximate the ``fixed point'' of this operator as a boundary MPS. One can verify that the only difference with the linear case is the presence of the corner, which can be captured by the insertion of a corner matrix $S$, which acts on the virtual level of the uniform MPS and is depicted with a diamond shape below. The fixed point equation then reduces to a linear equation in $S$ so that the optimal solution is easily found \footnote{We refer to the Supplemental Material for additional details on this contraction scheme.}.
\par Using the fixed points of all four corner transfer matrices, we can construct an effective environment consisting of different one-dimensional "channels", for which we can compute expectation values exactly using standard MPS techniques \cite{Schollwock2011a}. More specifically, with the channel structure we can compute two- and three-point functions for operators at arbitrary relative positions by contracting networks such as
\begin{equation} \label{eq:threePoint}
\figL{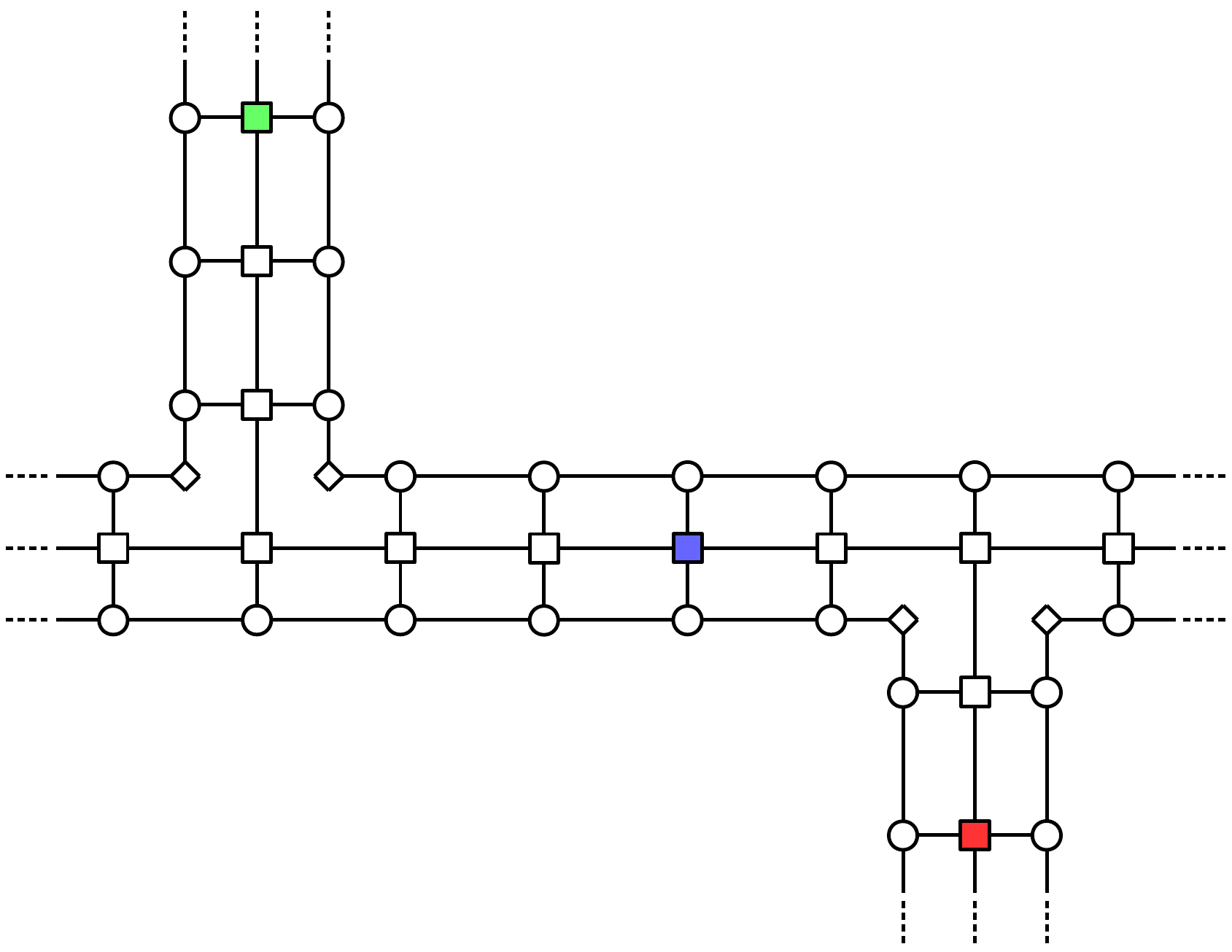}
\end{equation}
where the colored tensors indicate where the operators are located. In the Supplemental Material we show how to compute fourier transforms of these types of diagrams directly, thus enabling the evaluation of e.g. static structure factors and the matrix elements featuring in Eq.~\eqref{eq:rayleighritz}.

\paragraphL{Magnons in the AKLT model} As a first application, we study the two-dimensional AKLT model \cite{Affleck1987,*Affleck1988} on the square lattice. The one-dimensional AKLT model was first introduced to establish the existence of a gap in a rotationally invariant \mbox{spin-1} chain, in support of the Haldane conjecture \cite{Haldane1983, *Haldane1983a}. The construction can be straightforwardly extended to two-dimensional systems, but the existence of a gap -- although highly expected -- can no longer be established rigorously. On the square lattice the AKLT state can be represented as a PEPS with bond dimension $D=2$, and is the unique ground state of
\begin{equation*}
H_\text{AKLT} = \frac{1}{28}\sum_{\braket{ij}} h_{ij}  + \frac{7}{10}h_{ij}^2 + \frac{7}{45} h_{ij}^3 + \frac{1}{90} h_{ij}^4
\end{equation*}
with $h_{ij}=\vec{S}_i\cdot\vec{S}_j$ the spin-2 Heisenberg interaction.
\par
\begin{figure} 
\includegraphics[width=0.99\columnwidth]{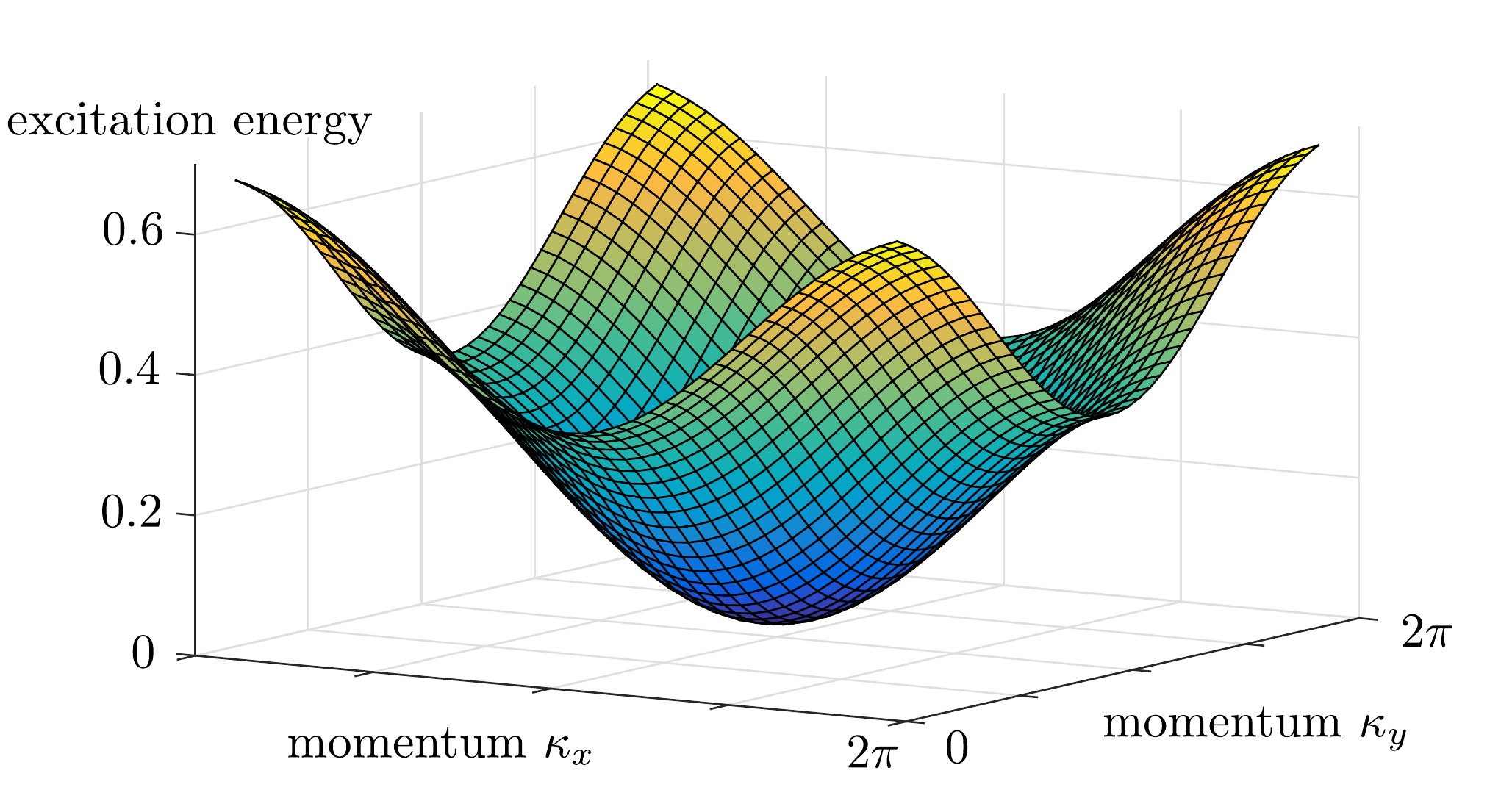}
\caption{Single mode approximation for the dispersion relation of the two-dimensional AKLT model on a square lattice. The minimum of the triplet dispersion relation is at momentum $\kappa_{x,y}=\pi$.}
\label{fig:sma} 
\end{figure}

 The elementary excitations in this model are expected to be (triply-degenerate) magnons. In a first step, we can target the magnons with the single-mode approximation (SMA), which is known to reproduce the one-particle dispersion relation qualitatively in the one-dimensional case \cite{Arovas1988}. In the present context the SMA wave function is given by ($\alpha=x,y,z$)
\begin{equation*}
\ket{\Phi_{\kappa_x,\kappa_y}^\alpha}_\text{SMA} = \sum_{m,n} \e^{i(\kappa_xm+\kappa_yn)} \hat{S}^\alpha_{(m,n)}\ket{\Psi_\text{AKLT}}.
\end{equation*}
The norm of these states $\braket{\Phi_{\kappa_x,\kappa_y}^\alpha|\Phi_{\kappa_x,\kappa_y}^\alpha}$ is equal to the static structure factor and can be computed with our new contraction scheme \footnote{The structure factor of the AKLT model can be shown to follow the 2D Ornstein-Zernike form \cite{Arovas1988}; in the Supplemental Material we show that our computations can reproduce this form accurately.} while the energy expectation value reduces to an easy PEPS contraction. The dispersion relation is shown in Fig.~\ref{fig:sma}; the  spectrum consists of an elementary triplet with its minimum at $\kappa_{x,y}=\pi$ \cite{Zauner2015} and a gap $\Delta_\text{SMA}=0.0199$.
\par Next we can determine the excitations variationally with the excitation ansatz [Eq.~\eqref{eq:ansatz}]. To improve on the SMA result, we enlarge our variational subspace by perturbing the ground state on a larger region -- a procedure that is bound to converge exponentially fast to the correct wave function \cite{Haegeman2013a}. By introducing a $B$ tensor on a block of two by two sites, we are able to estimate the gap at $\Delta_\text{var} = 0.0147$ \footnote{This value is quite small because of the prefactor of $\frac{1}{28}$ making the AKLT Hamiltonian a projector.}, in excellent agreement with the value $2\Delta=0.03$ in Ref. \onlinecite{Garcia-Saez2013}, obtained through a computation of the magnetization curve. In addition, we can compute the characteristic velocity through the second derivative of the dispersion relation in its minimum and we obtain $v_\text{var}=0.04115$. Finally, in Table~\ref{tab:weights} we have listed the spectral weight of the elementary magnon around its minimum.
\begin{table}
\centering
\begin{tabular}{|c|c|c|}
\hline
$\rho$ & spectral weight $w$ & \% of sum rule  \\ \hline
$0.0000\pi$ & $31.222$ & $99.57$ \\ \hline
$0.0682\pi$ & $26.058$ & $99.35$ \\ \hline
$0.1364\pi$ & $17.198$ & $98.84$ \\ \hline
$0.2045\pi$ & $10.672$ & $96.59$ \\ \hline
$0.2273\pi$ & $8.9968$ & $94.26$ \\ \hline
$0.2500\pi$ & $7.3452$ & $88.75$ \\ \hline
\end{tabular}
\caption{Spectral weight $w=\sum_\alpha\left|\bra{\Phi_\alpha}S^z\ket{\Psi(A)}\right|^2$ of the elementary magnon triplet as a function of momentum $\kappa_x=\kappa_y=\pi+\rho/\sqrt{2}$, and the percentage of the integrated spectral function that is saturated in the one-particle sector (we can compute the integrated spectral function exactly through the sum rule \cite{Hohenberg1974} and the static structure factor, see Supplemental Material). We observe that the magnon contains nearly all spectral weight in the minimum, and that this percentage drops when going away from this point.}
\label{tab:weights}
\end{table}

\paragraphL{Anyons in the perturbed toric code} As a second example we study the anyonic excitations in the toric code \cite{Kitaev2003}, the easiest example of a model exhibiting topological order and anyonic excitations (fluxes and charges). At the toric code fixed point, the anyons have a flat dispersion relation since all Hamiltonian terms commute. To generate non-trivial dynamics, we perturb the toric code state corresponding to a non-unitary operator
\begin{equation*}
f_i = \exp \left( \frac{1}{4}  (\beta_x \sigma_i^x + \beta_z\sigma_i^z) \right).
\end{equation*}
acting on every site $i$. The associated Hamiltonian is
\begin{align*}
H_\text{FTC} &= \sum_s\tilde{H}_s + \sum_p\tilde{H}_p
\end{align*}
where the filtered star operators are given by
\begin{align*}
\tilde{H}_{s} = \left( \prod_{i\in s} f^{-1}_i \right) \left( 1- \prod_{i} \sigma_i^x \right) \left( \prod_{i\in s} f^{-1}_i \right)
\end{align*}
and similar for the plaquette operator $\tilde{H}_{p}$ with $\sigma_i^z$ for $i\in p$. In first order, the filtering operation is equivalent to applying a magnetic field to the toric code. In fact, it has been shown in Refs. \onlinecite{Haegeman2015a,Haegeman2014b} that the phase diagram is qualitatively similar.
\par In a number of recent works it has been established that PEPS can provide a natural description of topological phases, where the topological order of the global state is reflected by a symmetry of the local PEPS tensor $A$ on the virtual level \cite{Schuch2010a, Sahinoglu2014, *Buerschaper2014}. A variational ansatz for the complete set of elementary excitations in the different anyon sectors is obtained by attaching a half-infinite virtual matrix product operator string to the local $B$ tensor in the excitation ansatz [Eq.~\eqref{eq:ansatz}]. The topological sector is encoded in the type of string and the virtual symmetry representation of the local tensor. In the case of the toric code, the flux of the excitations is encoded in the presence or absence of a string of $\sigma^z$ operators, whereas the charge is encoded in the symmetry representation of the $B$ tensor \cite{Schuch2010a}. These topological characteristics are unaltered by the filtering.
\par
\begin{figure}
\includegraphics[width=0.99\columnwidth]{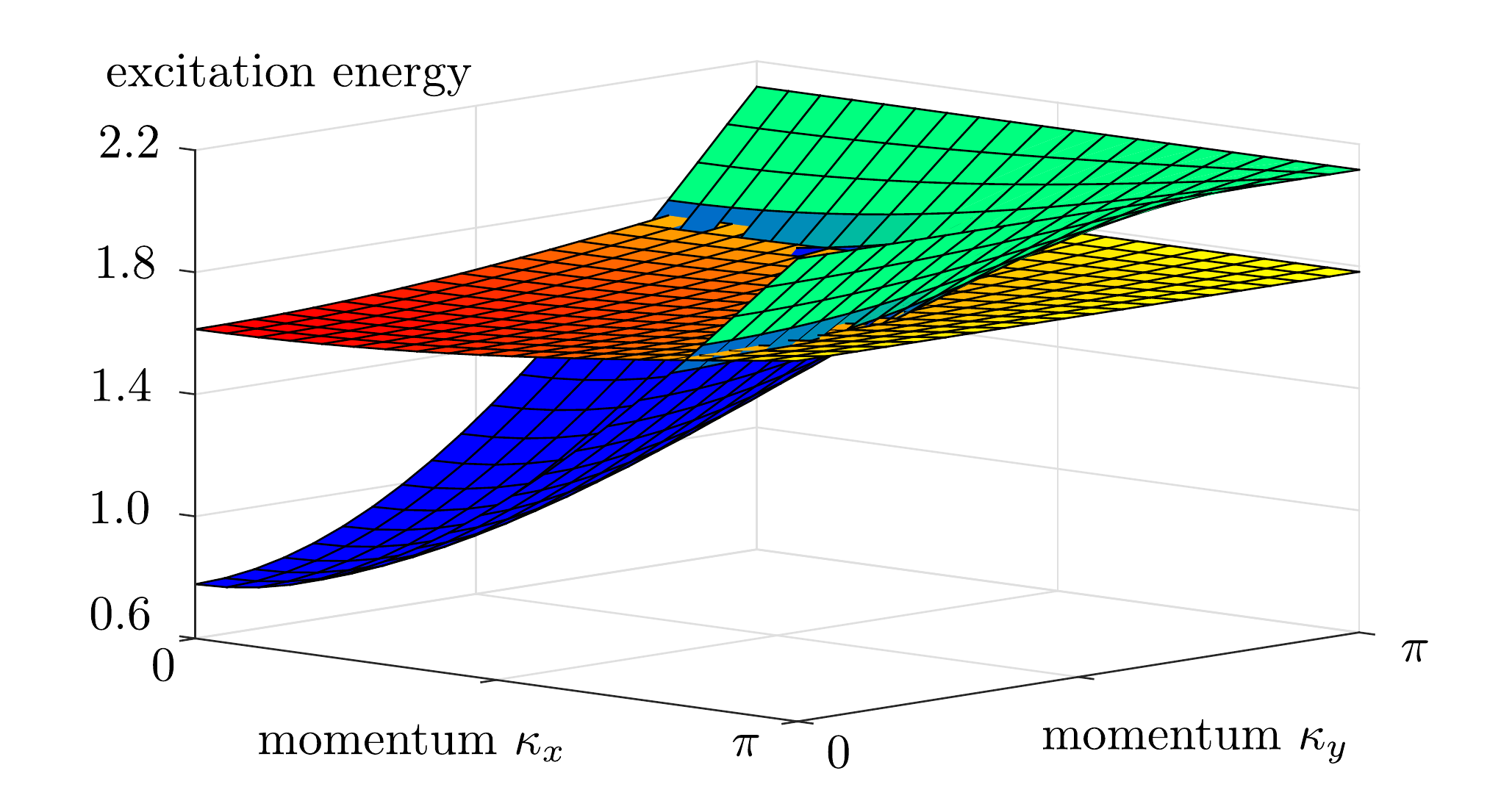}
\caption{The dispersion relation of the toric code in the flux (blue-green) and charge (red-yellow) sector for filtering parameters $\beta_x=0.05$ and $\beta_z=0.35$. For the former, we have used the topological ansatz with a string of $\sigma^z$ operators at the virtual level, whereas the latter was obtained with an excitation that carries no string but transforms according to the odd representation of the virtual $\mathbb{Z}_2$ symmetry (note that we did not impose this symmetry on the $B$-tensor in Eq.~\eqref{eq:ansatz} explicitly).}
\label{fig:dispToric} 
\end{figure}

\begin{figure}
\includegraphics[width=0.99\columnwidth]{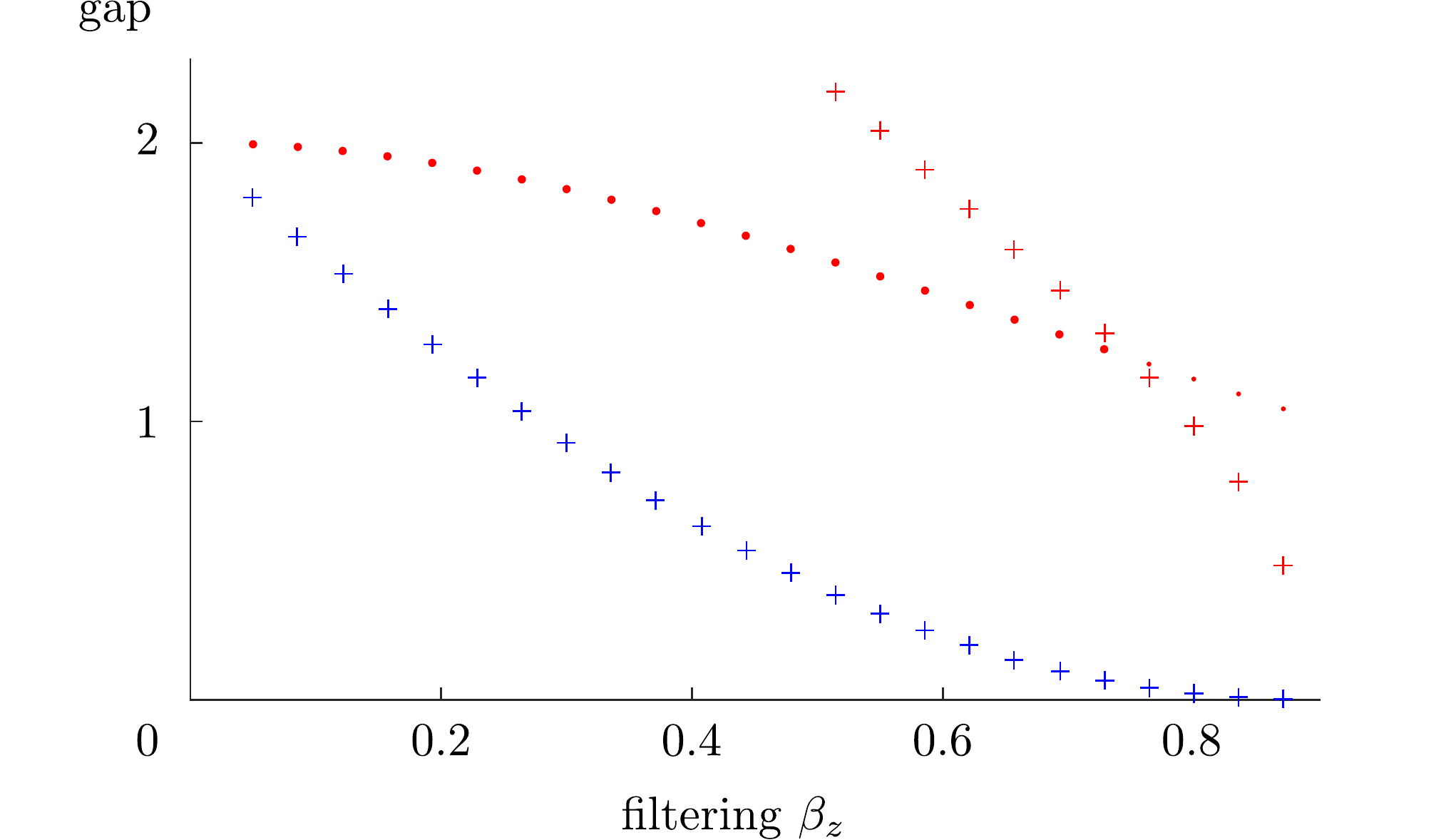}
\caption{The lowest-lying excitation energies in the charge and flux sector as a function of $\beta_z$ and with $\beta_x=0$, for which the model can be shown to have a phase transition at $\beta_z=\log(\sqrt{2}+1)\approx0.88$ \cite{Castelnovo2008}. The color represents trivial (red) or non-trivial (blue) flux of the excitations, whereas the symbol represents even ($+$) or odd ($\cdot$) charge number. We can observe that the gap of the flux excitation (even charge in the non-trivial sector) closes at the phase transition, signalling the condensation of the flux excitations. We can even see the two-flux state (even charge in the trivial sector) coming down in energy close to the phase transition, although our variational ansatz is not suited for describing multi-particle states. The charge excitation (odd charge in the trivial sector) remains gapped.}
\label{fig:gapToric} 
\end{figure}

 In Fig.~\ref{fig:dispToric} we have plotted the elementary excitation spectrum in both the charge and flux sector for a specific value of the filtering parameters. We can see that the filtering in one direction mostly affects one of the two excitations. This is reflected more clearly in Fig.~\ref{fig:gapToric}, where we have plotted the gap in both sectors along the $\beta_x=0$ axis. At the critical point $\beta_z=\log(\sqrt{2}+1)$, one can observe that the gap to the one-flux and many-flux states closes, indicating the condensation of flux excitations. The charge sector remains gapped but will cease to exist beyond the transition, for reasons explained in Ref.~\onlinecite{Haegeman2014b}.

\paragraphL{Conclusions}
This paper introduced the concept of post-PEPS methods as a systematic approach for studying the low-energy spectrum of two-dimensional quantum systems, as well as a new PEPS contraction scheme that enables to compute the necessary quantities. 
\par In the one-dimensional case, the local description of excitations in the MPS framework has led to the development of an effective particle picture of excitations on top of a strongly-correlated vacuum \cite{Vanderstraeten2014, *Vanderstraeten2015a}. Our methods open up the route to a two-dimensional generalization of this particle description of the low-energy dynamics and, more specifically, to the formulation of the scattering problem for effective particles in two dimensions. Applications such as multi-particle spectral properties, bound-state formation, magnon condensation and low-temperature behaviour seem within reach.
\par We have also introduced a new contraction scheme, which will allow for PEPS algorithms based on the time-dependent variational principle \cite{Dirac1930, *Frenkel1934, Haegeman2011d}. These will prove useful to simulate the real-time dynamics of quantum quenches, but might also result in improved algorithms for finding PEPS ground state approximations. Indeed, with our new effective environment we are able to compute both the gradient and the Hessian of the (global) energy functional, which can be used as an input for numerical optimization methods \cite{preparation}.

\begin{acknowledgments}
Research supported by the Research Foundation Flanders (LV, MM, JH), the Austrian FWF SFB grants FoQuS and ViCoM, and the European grants SIQS and QUTE (FV).
\end{acknowledgments}

\bibliography{bibliography}

\onecolumngrid
\appendix

\section*{SUPPLEMENTAL MATERIAL}

\paragraphL{\textbf{Transfer matrix and fixed points}} In the first part of this Supplemental Material we study the fixed points of the linear transfer matrix $\mathcal{T}$ and corner transfer matrix $\mathcal{C}$. For a translation-invariant MPS to satisfy the fixed point equation of the linear transfer matrix $\mathcal{T}$
\begin{equation*}
\figL{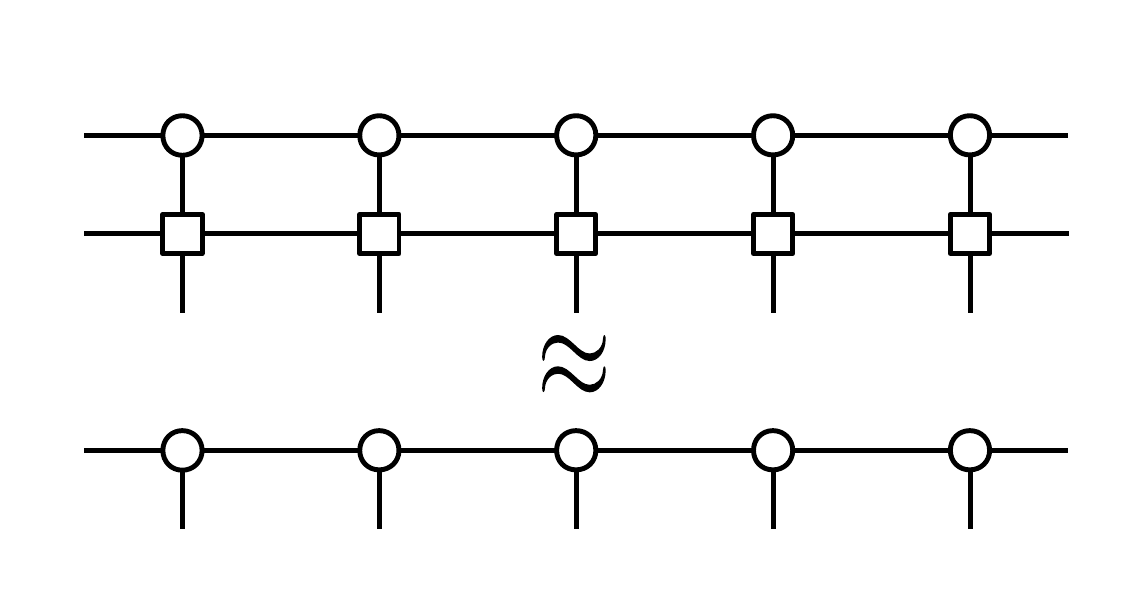}
\end{equation*}
it is required that there exists a tensor (rectangle) such that
\begin{equation} \label{eq:rectangletensor}
\figL{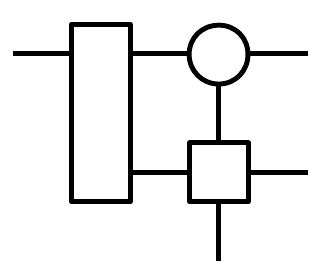} \approx \figL{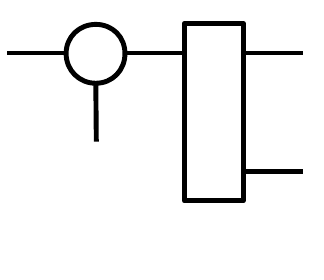}
\end{equation}
holds up to small truncation errors.
\par Let us now show that the ``fixed point'' of the corner transfer matrix $\mathcal{C}$ can indeed be approximated by introducing a single corner tensor $S$ between the MPS tensors of the fixed points of respectively the horizontal and vertical linear transfer matrix $\mathcal{T}$; pictorially, we have
\begin{equation*}
\figL{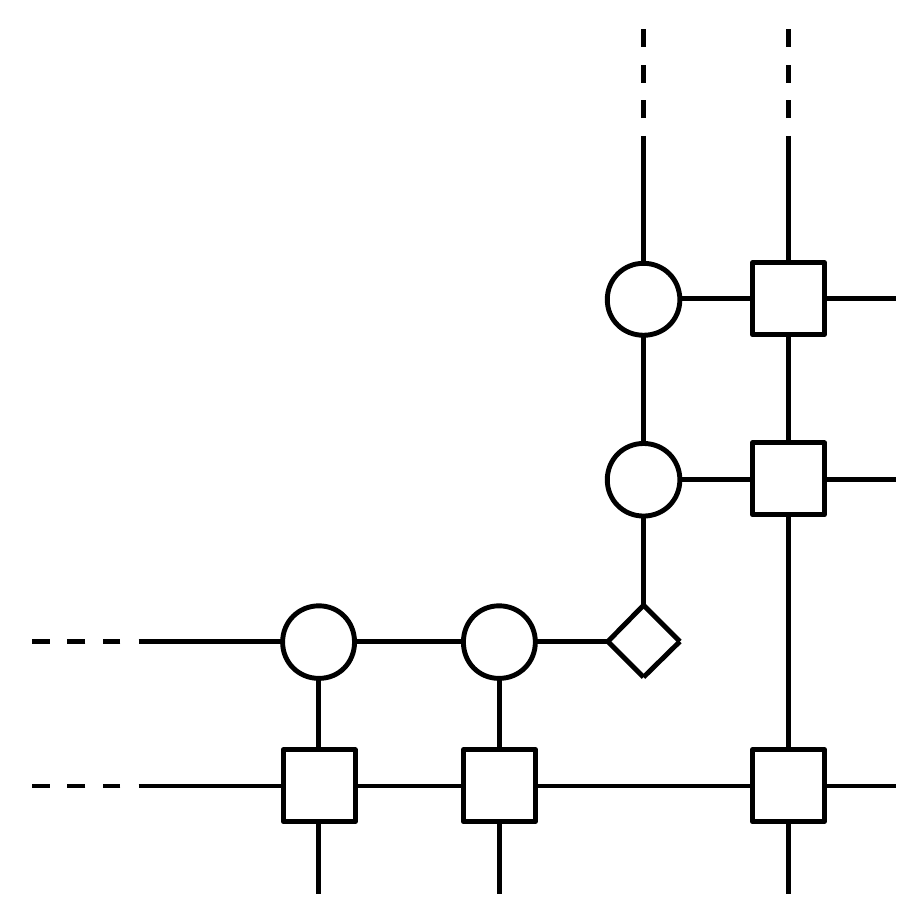} \propto \figL{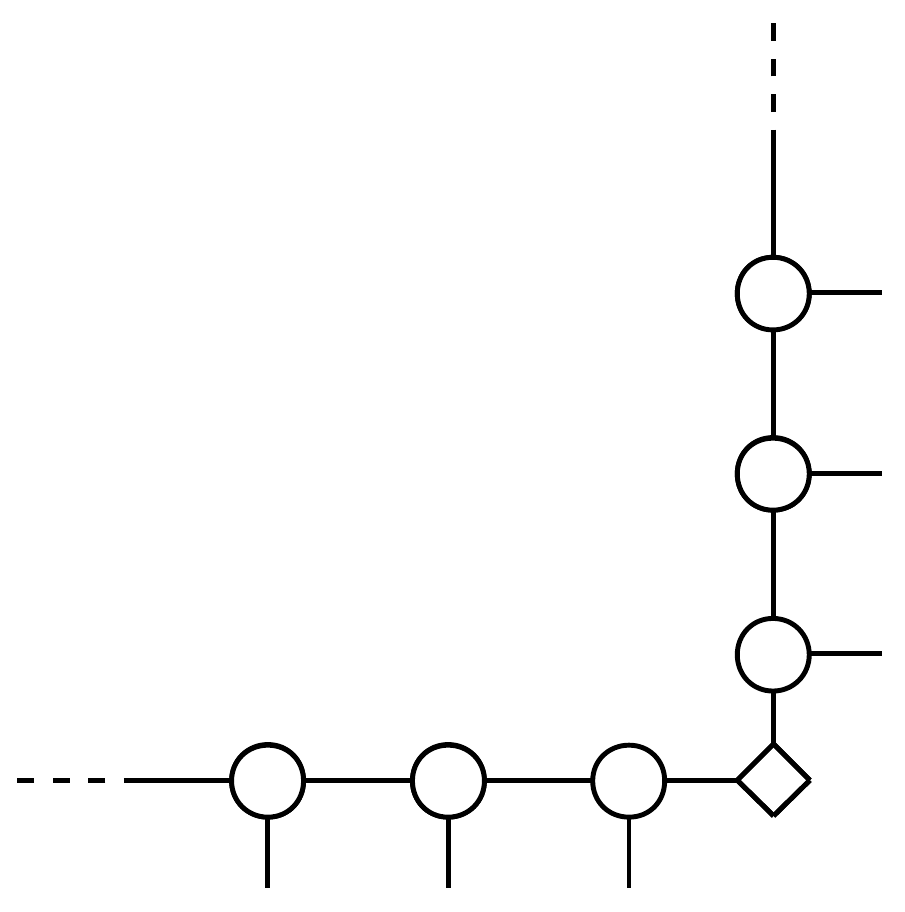}.
\end{equation*}
Note, firstly, that the corner transfer matrix introduces two new sites at every application (similar to infinite-size DMRG algorithms \cite{White1992, McCulloch2008}) and therefore does not have a fixed point in the strict sense. Nevertheless, repeated application of $\mathcal{C}$ is likewise expected to result in a state with a converged (i.e. translation-invariant) structure, up to the corner itself. We therefore model the ``fixed point'' using the tensors of the fixed point of the linear transfer matrix and insert a new corner tensor. Applying the corner transfer matrix $\mathcal{C}$ once and using the tensor from Eq.~\eqref{eq:rectangletensor} gives rise to
\begin{equation*}
\figL{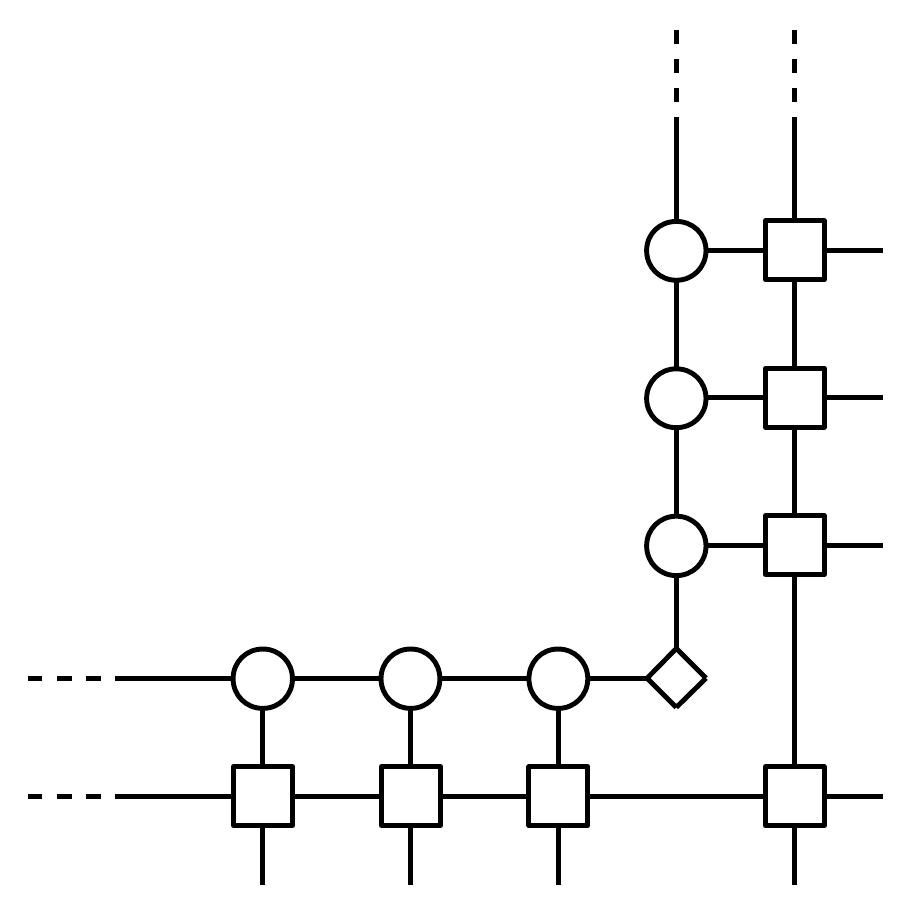} \approx \figL{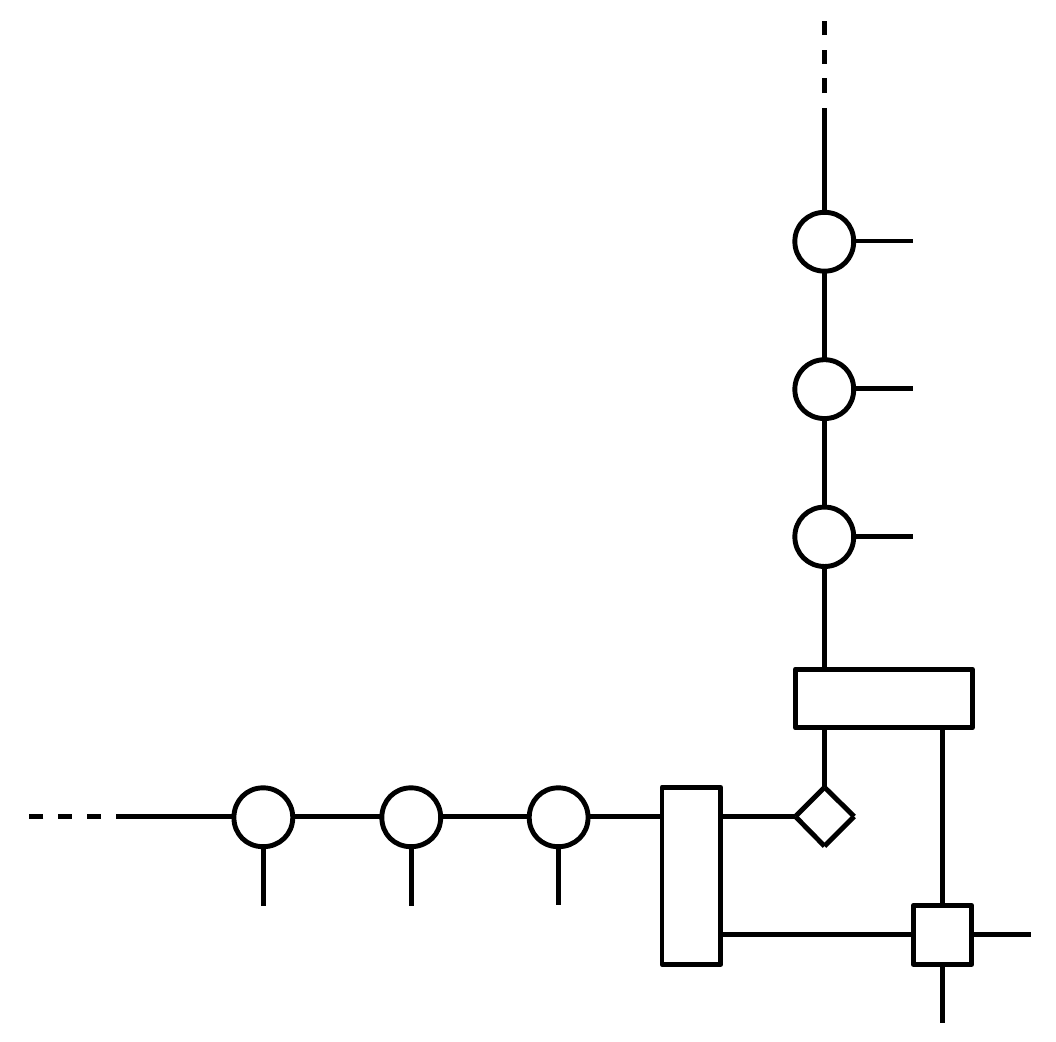},
\end{equation*}
which shows that the original MPS tensors are indeed obtained after application of $\mathcal{C}$, except on the two newly introduced sites. In principle, two new MPS tensors connected by a new corner matrix could appear. However, since the unique MPS tensor of the fixed point of $\mathcal{T}$ seem to capture the correct structure on the further sites, and these two can be assumed to have originated from previous applications of $\mathcal{C}$, we can make the ansatz that these tensors should also be put on the two new sites. With this ansatz, we obtain a linear fixed point equation for the corner matrix itself, which corresponds to a simple eigenvalue equation. With the corner matrix as only variational parameters in the fixed point equation for $\mathcal{C}$, we essentially have a linear subspace as ansatz and can therefore easily measure the error obtained by projecting onto this subspace.

\paragraphL{\textbf{Excitation ansatz: some details}} The variational ansatz for an elementary excitation is given by
\begin{equation*}
\ket{\Phi_{\kappa_x\kappa_y}(B)} = \sum_{m,n} \e^{i(\kappa_xm+\kappa_yn)} \\ \figL{ansatz}.
\end{equation*}
The blue tensor $B$, containing all variational freedom in the ansatz, has the same dimensions as the ground state tensor $A$, yet the number of variational parameters is smaller because of a redundancy in the parametrization. Indeed, through simple insertion one can easily check that the tensors (where the green square is a $D\times D$ matrix)
\begin{equation*}
B_{0,x} = \figL{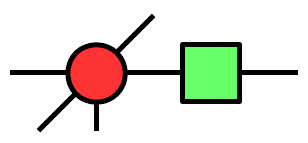} - \e^{i\kappa_x} \figL{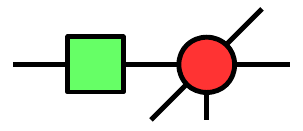}
\end{equation*}
and
\begin{equation*}
B_{0,y} = \figL{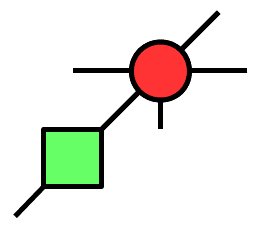} - \e^{i\kappa_y} \figL{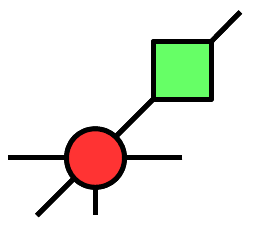}
\end{equation*}
give rise to states with zero norm. Correspondingly, the \emph{effective norm matrix} $\mathsf{N}_\text{eff}^{\kappa_x\kappa_y}$, defined as
\begin{equation*}
\braket{\Phi_{\kappa'_x\kappa'_y}(B')|\Phi_{\kappa_x\kappa_y}(B)} = 4\pi^2\delta(\kappa_x-\kappa'_x)\delta(\kappa_y-\kappa'_y) \; \vect{B}' \, \mathsf{N}_\text{eff}^{\kappa_x\kappa_y} \, \vect{B},
\end{equation*}
will have a number of zero eigenvalues ($\vect{B}$ is the vector containing all elements of the tensor $B$). In order for the variational subspace to be well-defined, we will always project out these null modes. In addition, since we want an excitation to be orthogonal to the ground state, we will also project out the component that is proportional to the ground state tensor $A$. 
\par Once we have defined the variational subspace, we can minimize the energy in order to find the best approximation to the true excitation. Since the subspace is linear, this can be done by solving the Rayleigh-Ritz problem
\begin{equation*}
\mathsf{H}_\text{eff}^{\kappa_x\kappa_y} \vect{B} = \omega \mathsf{N}_\text{eff}^{\kappa_x\kappa_y} \vect{B}
\end{equation*} 
where the \emph{effective Hamiltonian matrix} $\mathsf{H}_\text{eff}^{\kappa_x,\kappa_y}$ is similarly defined as ($E_0$ is the ground-state energy)
\begin{equation*}
\bra{\Phi_{\kappa'_x\kappa'_y}(B')}\hat{H}-E_0\ket{\Phi_{\kappa_x\kappa_y}(B)} = 4\pi^2\delta(\kappa_x-\kappa'_x)\delta(\kappa_y-\kappa'_y) \; \vect{B}' \, \mathsf{H}_\text{eff}^{\kappa_x\kappa_y} \, \vect{B}.
\end{equation*}
The matrix elements of $\mathsf{N}_\text{eff}^{\kappa_x,\kappa_y}$ and $\mathsf{H}_\text{eff}^{\kappa_x,\kappa_y}$ contain two- and three-point functions that can be computed with a channel environment as explained in the main body. In the remainder of this Appendix we show how to do this in detail.

\paragraphL{\textbf{Two-point functions}} We will first look at the effective norm matrix, which yields the following double (infinite) sum
\begin{equation*}
\braket{\Phi_{\vec{\kappa}'}[B']|\Phi_{\vec{\kappa}}[B]} = \sum_{\vec{n},\vec{n}'} \e^{i(\vec{\kappa}\cdot\vec{n} - \vec{\kappa}'\cdot\vec{n}')} \left[ \text{$B$ at site $\vec{n}$ and $B'$ at site $\vec{n}'$} \right].
\end{equation*}
Because the ground state is translation invariant, we can simplify to a single sum as
\begin{equation*}
\braket{\Phi_{\vec{\kappa}'}[B']|\Phi_{\vec{\kappa}}[B]} = 4\pi^2 \delta^2(\vec{\kappa}'-\vec{\kappa}) \sum_{m,n'} \e^{i\kappa_Vm}\e^{-i\kappa_Hn'} \left[ \text{$B$ at site $(m,0)$ and $B'$ at site $(0,n')$} \right] .
\end{equation*}
These terms are all two-point functions, which can be computed with our channel environment. One term looks like
\begin{equation*}
\figL{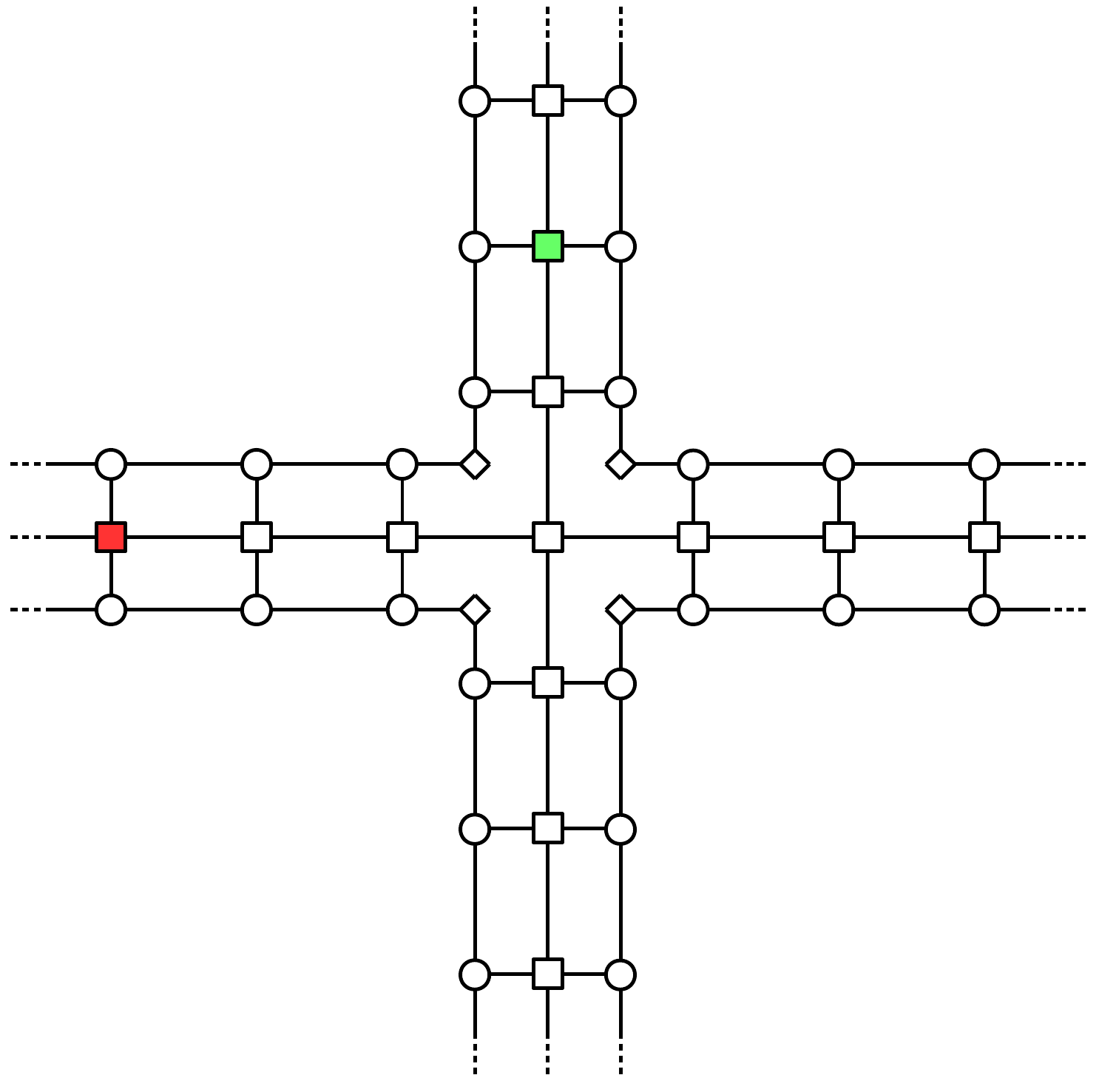}
\end{equation*}
where the green (red) tensors indicate the locations of $B$ and $B'$. In a first step we can contract the infinite channels by computing the fixed point of the ``channel operators'' and represent these as three-legged rectangles
\begin{equation*}
\figL{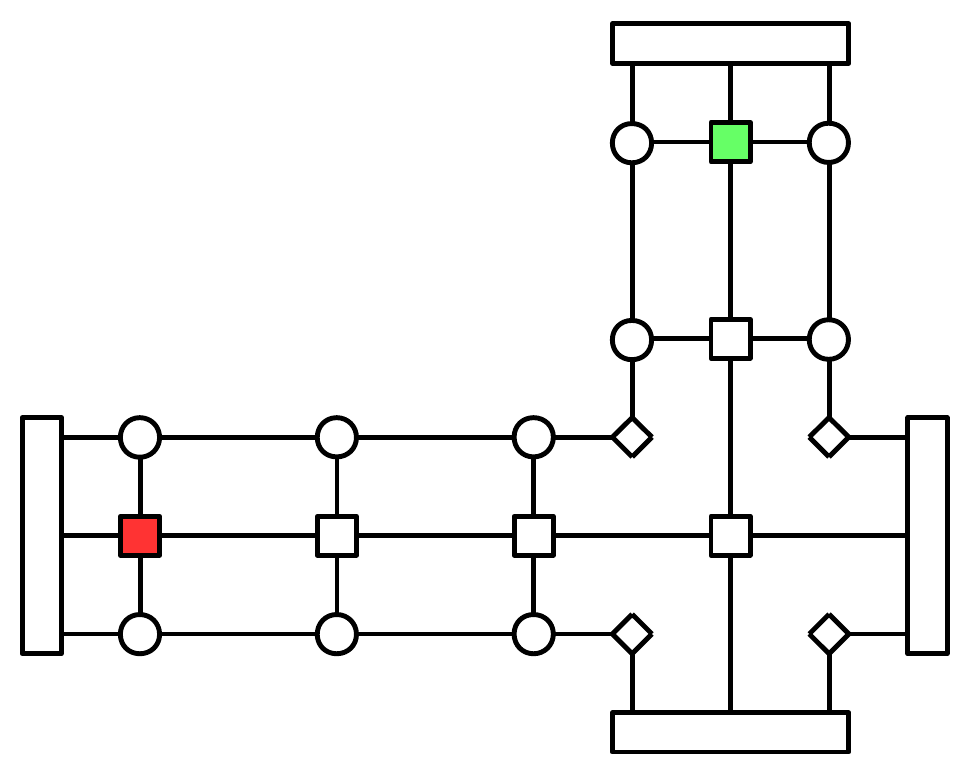}.
\end{equation*}
The momentum superpositions can be worked out by inverting the relevant channel operator. For example, we can write one of the sums as
\begin{align*}
& \sum_{m=1}^{\infty} \e^{i\kappa_Vm}\ \left[ \text{$B$ at site $(m,0)$ and $B'$ at site $(0,0)$} \right] \\
& = \e^{i\kappa_V} \figL{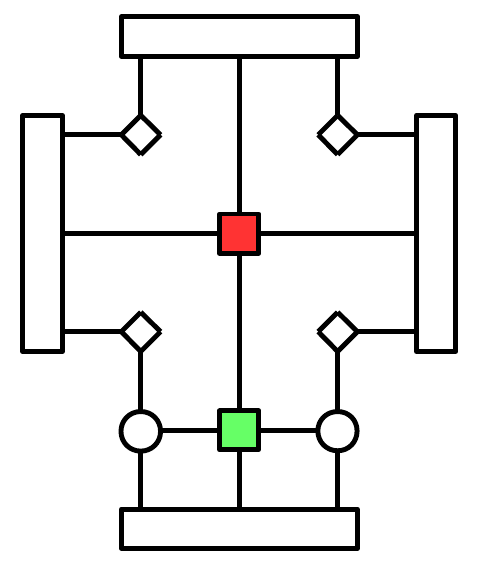} + \e^{2i\kappa_V} \figL{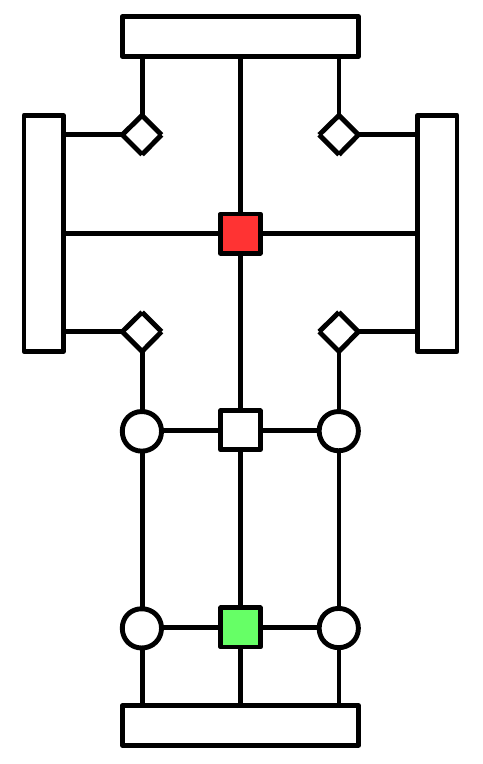} + \e^{3i\kappa_V} \figL{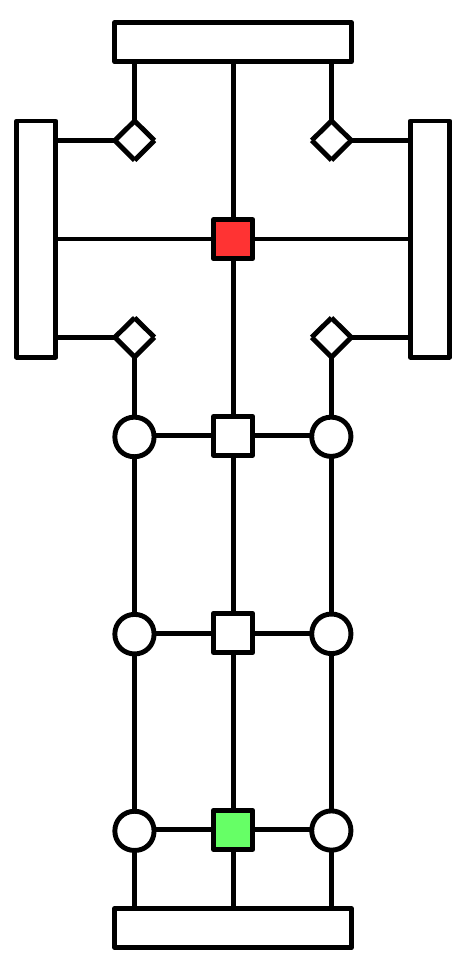} + \dots \\
&= \e^{+i\kappa_V} \figL{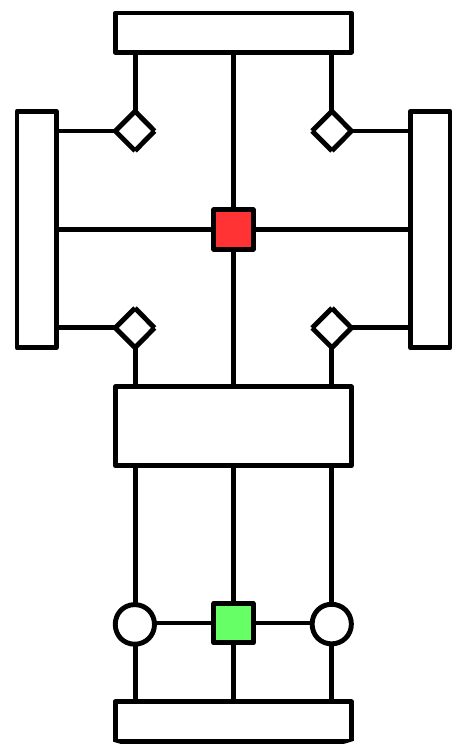}
\end{align*}
where
\begin{align} \label{inverse}
\figL{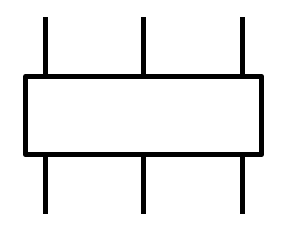} = \sum_{m=0}^{\infty} \e^{i\kappa m} \left( \figL{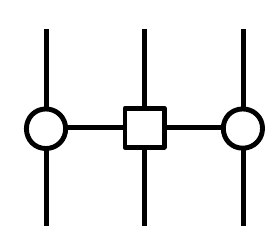} \right)^m = \left( 1-\e^{i\kappa} \figL{b4} \right) ^{-1}.
\end{align}
Applying this procedure to all geometric sums and keeping the local term amounts to
\begin{align}
& \braket{\Phi_{\vec{\kappa}'}[B']|\Phi_{\vec{\kappa}}[B]} = 4\pi^2 \delta^2(\vec{\kappa}'-\vec{\kappa}) \times \Bigg[ \figL{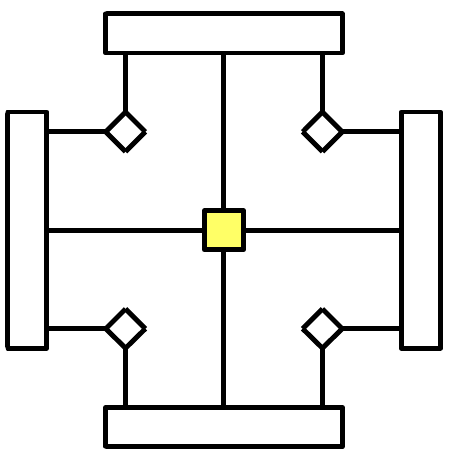} \nonumber \\
& \qquad + \e^{-i\kappa_V} \figL{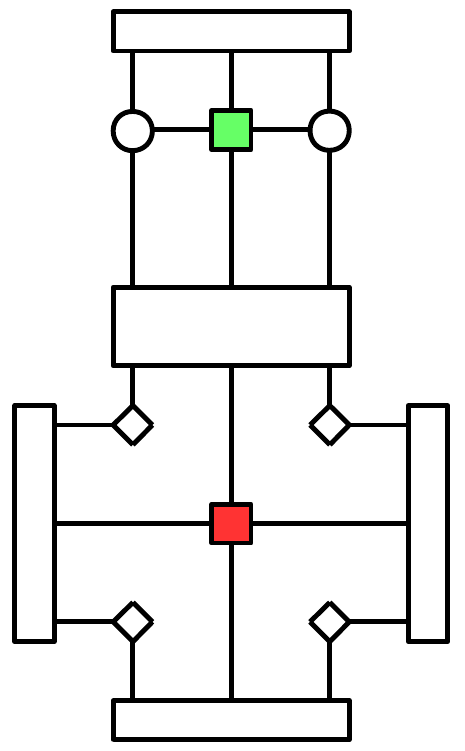} + \e^{+i\kappa_V} \figL{c3} \nonumber \\
& \qquad + \e^{+i\kappa_H} \figL{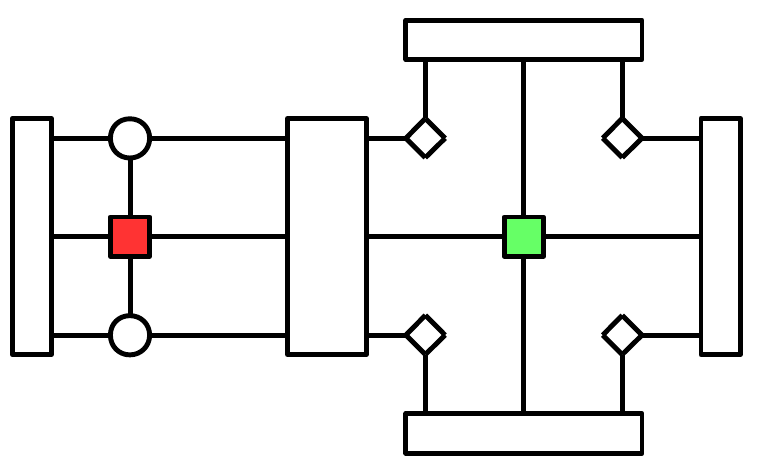} + \e^{-i\kappa_H} \figL{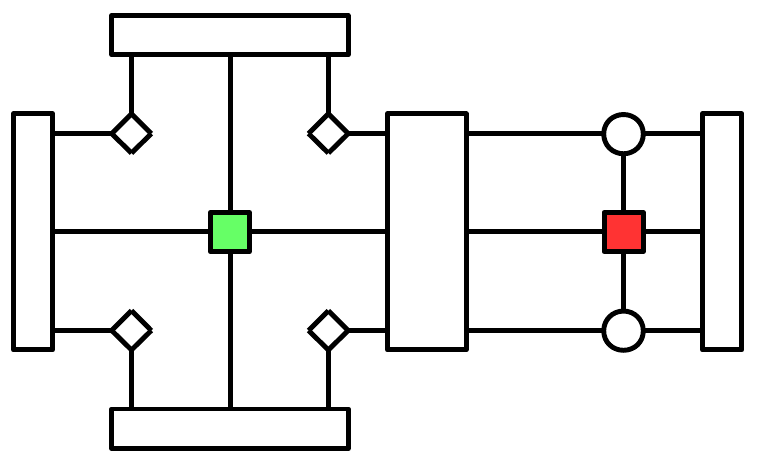} \nonumber \\
& \qquad + \e^{-i\kappa_V}\e^{+i\kappa_H} \figL{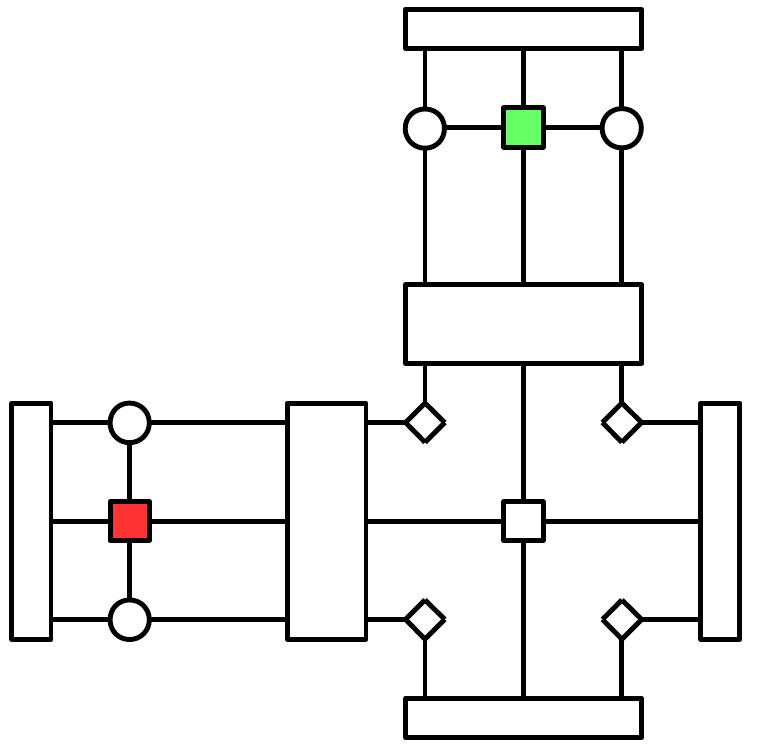} + \e^{-i\kappa_V}\e^{-i\kappa_H} \figL{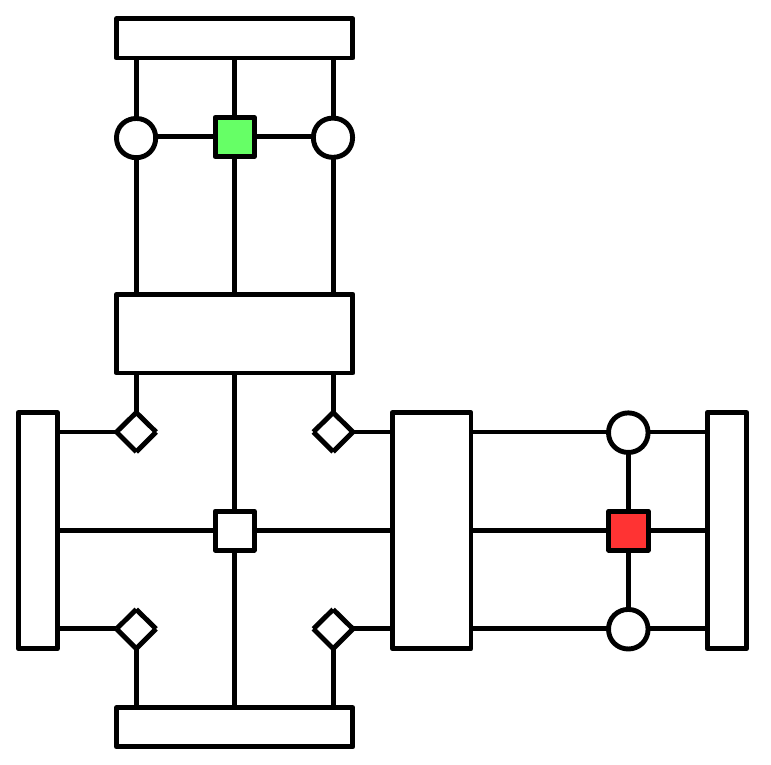} \nonumber \\
& \qquad + \e^{+i\kappa_V}\e^{+i\kappa_H} \figL{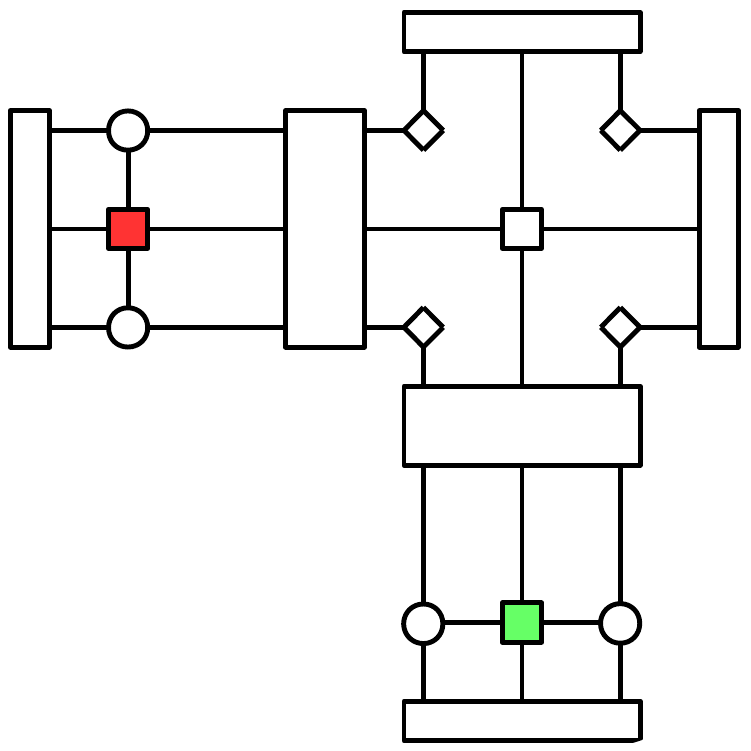} +\e^{+i\kappa_V}\e^{-i\kappa_H} \figL{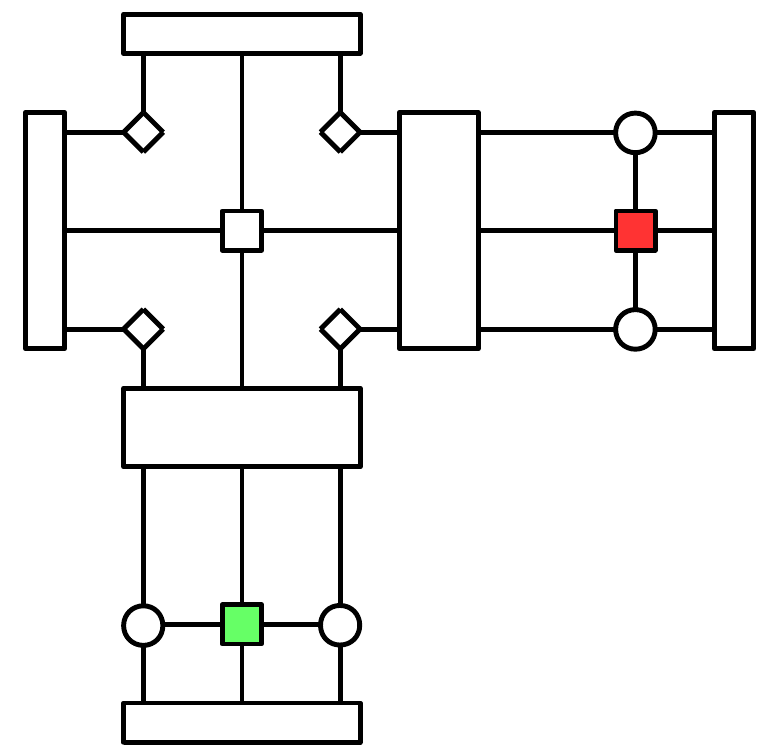} \Bigg], \label{eq:normA}
\end{align}
where a green (red) square indicates the transfer operator with a $B$ ($B'$) on the ket (bra) level and the yellow square represents the situation where $B$ and $B'$ are on the same site; the big six-leg tensors represent inverses as in \eqref{inverse}, the three-leg tensors are the fixed points of the channels and the diamond tensors are the corner matrices.
\par The overlap of the Hamiltonian with respect to two excitations can be calculated easily for a frustration-free Hamiltonian: disconnected terms are automatically zero as the Hamiltonian annihilates the ground state locally. The overlap $\bra{\Phi_{\vec{\kappa}'}[B']}\hat{H}\ket{\Phi_{\vec{\kappa}}[B]}$ reduces to a number of local contractions, which we will not write down (as they depend on the form of the Hamiltonian). 
\par Note that the non-frustration free case can be calculated equally well with our methods -- provided we have a good PEPS representation of the ground state. Disconnected terms will generally be of the form
\begin{equation*}
\figL{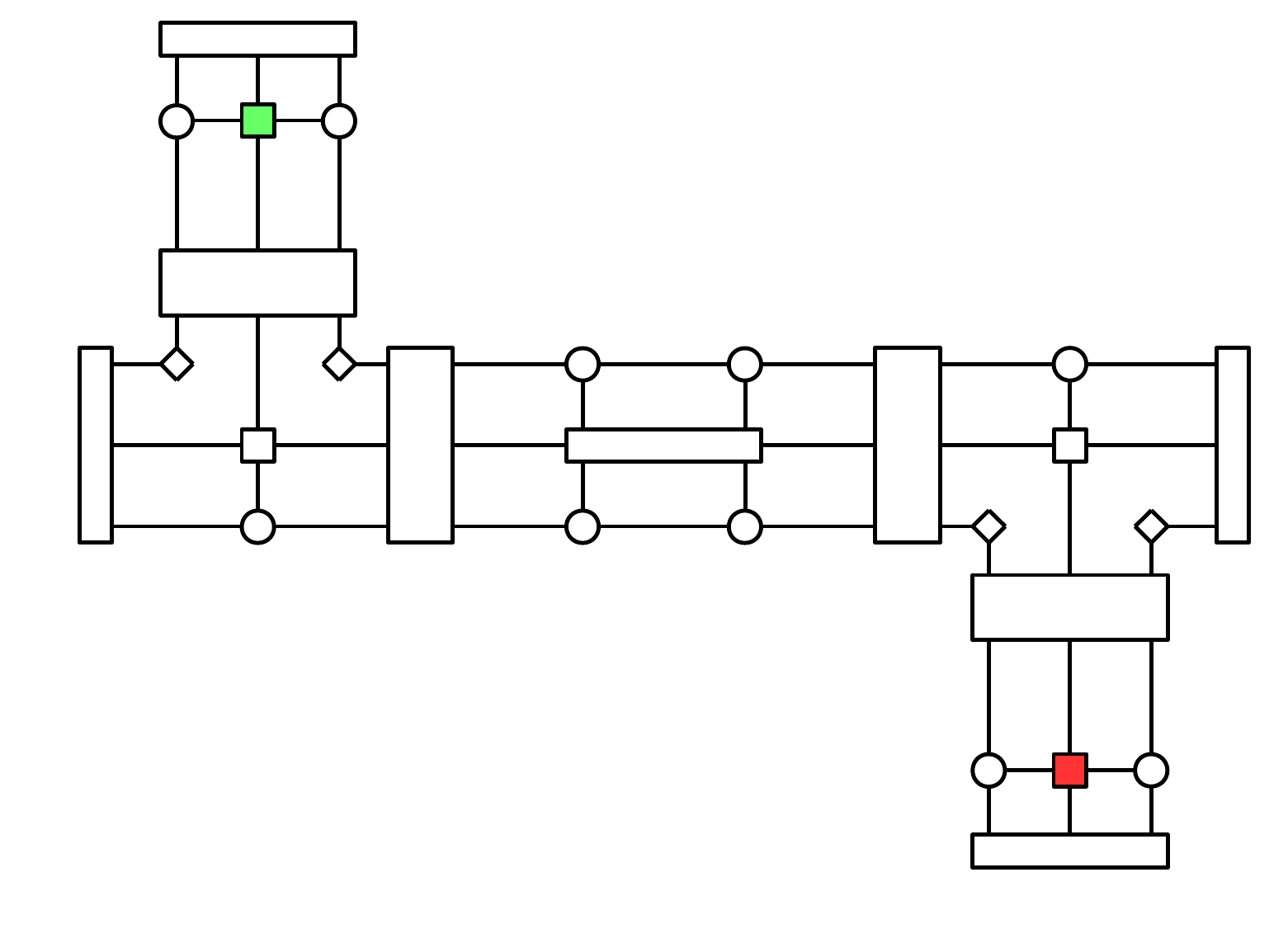}
\end{equation*}
where the middle six-leg tensor represents a two-site Hamiltonian term squeezed between ground state PEPS tensors.

\begin{figure} 
\includegraphics[width=0.6\columnwidth]{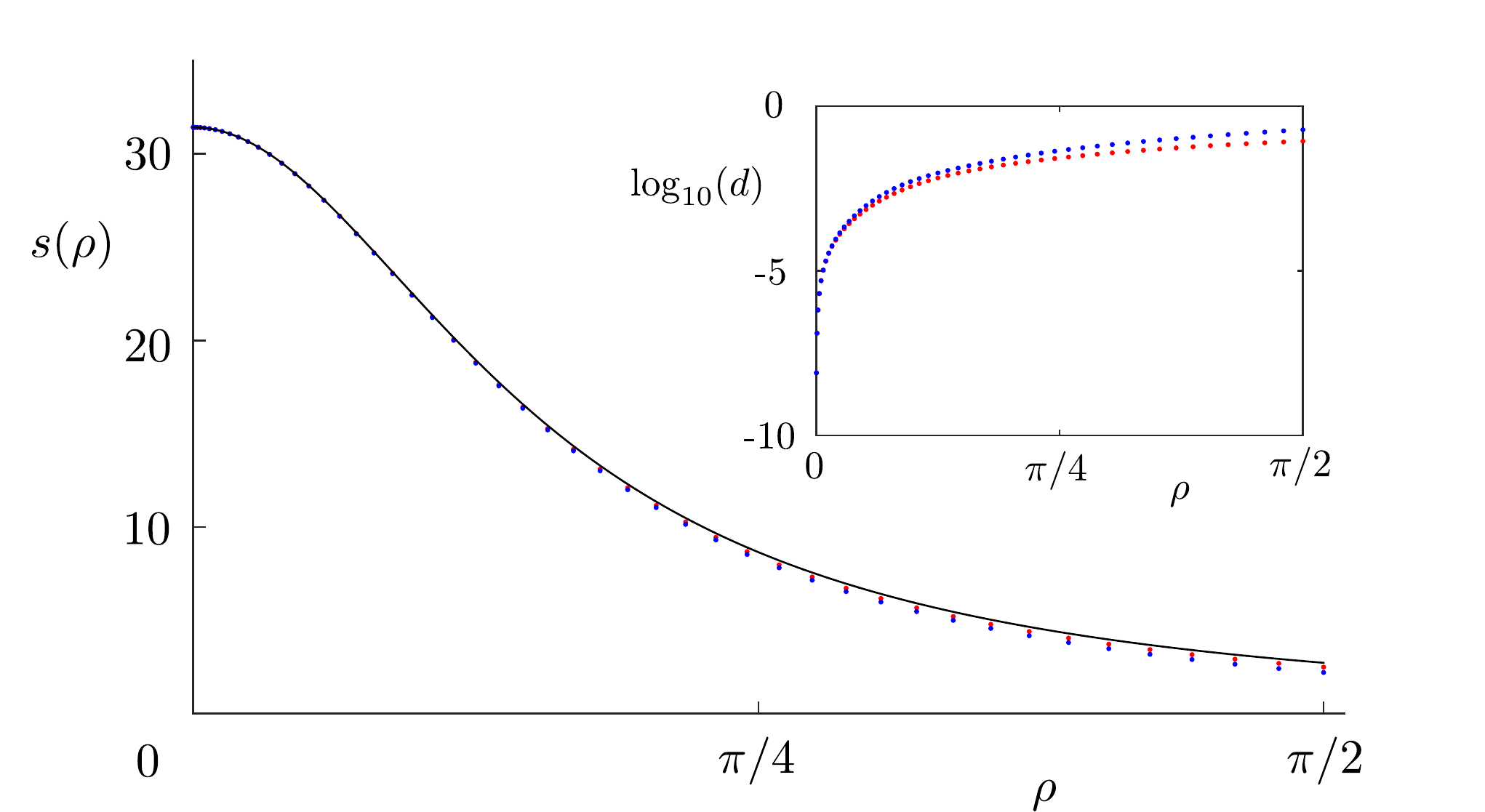}
\caption{The structure factor $s(\rho)$ where $\rho=((\kappa_x-\pi)^2+(\kappa_y-\pi)^2)^{1/2}$ for the two-dimensional AKLT model along the axes $\kappa_y=\pi$ (red) and $\kappa_x=\kappa_y$ (blue) compared to the form in Eq.~\eqref{eq:OZ}. In the inset the ($\log_{10}$ of the) deviations are plotted. The plot shows that the classical Ornstein-Zernike form is accurate for a large portion of the Brillouin zone and that the structure factor is nicely isotropic around $\vec\kappa^*=(\pi,\pi)$.}
\label{fig:structureFactor}
\end{figure}

\paragraphL{\textbf{Structure factor and sum rules in the AKLT model}} Let us first study the static correlations of the AKLT state. In Refs.~\onlinecite{Arovas1988, Arovas1992} it was argued that these correlations can be expressed as thermal averages of a related classical model in the same number of dimensions, contrary what one expects for generic quantum ground states. The real-space correlations then decay according to the 2D Ornstein-Zernike form
\begin{equation*}
s(\vec{n})=\bra{\Psi(A)} S^\alpha_{\vec{n}} S^\alpha_{\vec{0}} \ket{\Psi(A)}\propto\e^{-\lvert \vec{n}\rvert/\xi}\,\e^{i\vec{\kappa}^*\cdot\vec{n}}/ \sqrt{n} \qquad\qquad (n\gg1)
\end{equation*}
with $S^\alpha_{\vec{n}}(t)$ the spin operator at site $\vec{n}$ and $\vec{\kappa}^*$ the oscillation period. In momentum space, the structure factor is thus given by
\begin{align} 
s(\vec{\kappa}) &= \sum_{\vec{n}} \e^{-i\vec{\kappa}\cdot\vec{n}} s(\vec{n}) \nonumber \\
&= \bra{\Psi_0} S_{-\vec{\kappa}}^{\alpha\dagger} S^\alpha_0(0) \ket{\Psi_0} \nonumber \\
& \propto \frac{1}{1+\xi^2|\vec\kappa-\vec\kappa^*|^2} \label{eq:OZ}
\end{align}
for momenta close to $\vec{\kappa}^*$ where $s(\vec{\kappa})$ reaches its maximum (we have defined the momentum spin operator $S^\alpha_{\vec{\kappa}}=\sum_{\vec{n}}\e^{i\vec{\kappa}\cdot\vec{n}}S^\alpha_{\vec{n}}$). We can confirm this result with our methods. Firstly, we can calculate the correlation length by computing the gap of the linear transfer matrix $\mathcal{T}$ using the methods of Ref. \cite{Haegeman2014b}. We obtain the value $\xi_\text{AKLT} = -1/\log|\lambda^{(2)}_\mathcal{T}| = 2.06491$, in reasonable agreement with the value of $\xi^{-1}\approx0.52$ in Ref.~\onlinecite{Hieida1999}. In Fig.~\ref{fig:structureFactor} we compare the computation of the structure factor with our methods [Eq.~\eqref{eq:normA}] with the form of Eq.~\eqref{eq:OZ} and observe very good agreement in a large portion of the Brillouin zone.
\par The structure factor also shows up in the integrated spectral function. The latter is defined as
\begin{equation*}
S(\vec{\kappa},\omega) = \sum_{\vec{n}} \int \d t \, \e^{i(\omega t - \vec{\kappa}\cdot \vec{n})} \bra{\Psi_0} S_{\vec{n}}^{\alpha\dagger}(t) S^\alpha_0(0) \ket{\Psi_0}
\end{equation*}
with $S^\alpha_{\vec{n}}(t)$ the spin operator at site $\vec{n}$ in the Heisenberg picture. By inserting a resolution of the identity consisting of all excited states $\one=\sum_\gamma\ket{\gamma}\bra{\gamma}$ and only taking into account the one-particle states, we get the one-particle contribution to the spectral function
\begin{equation*}
S_\text{1p}(\vec{\kappa},\omega) = \sum_{\gamma\in\Gamma(\vec{\kappa})} \left|\bra{\gamma(\vec{\kappa})} S^\alpha_0 \ket{\Psi(A)} \right|^2
\end{equation*}
where $\Gamma(\vec{\kappa})$ is the set of all one-particle states $\ket{\gamma(\vec{\kappa})}$ with momentum $\vec{\kappa}$. Upon integrating the spectral function, we get the following sum rule \cite{Hohenberg1974}
\begin{align*}
\int \frac{\d \omega}{2\pi} S(\vec{\kappa},\omega) &= \int \frac{\d \omega}{2\pi} \bra{\Psi_0} S_{-\vec{\kappa}}^{\alpha\dagger} 2\pi\delta(\omega-\hat{H}) S^\alpha_0(0) \ket{\Psi_0} \\
&= \bra{\Psi_0} S_{-\vec{\kappa}}^{\alpha\dagger} S^\alpha_0(0) \ket{\Psi_0} = s(\vec{\kappa}).
\end{align*}

\end{document}

%% file: preamble.tex
\usepackage{graphicx}
\usepackage{graphics}
\usepackage{amsmath}
\usepackage{amssymb}
\usepackage{amsfonts}
\usepackage{dsfont}
\usepackage{braket}
\usepackage{color}
\usepackage{braket,slashed}
\usepackage[mathscr]{euscript}
\usepackage{hyperref}
\usepackage{subfigure}
\usepackage{xfrac}
\usepackage{bm}
\usepackage{pdfpages}

\allowdisplaybreaks[1]

\newcommand{\one}{\mathds{1}}

\renewcommand{\d}{\ensuremath{\mathrm{d}}}
\newcommand{\e}{\ensuremath{\mathrm{e}}}

\newcommand{\vect}[1]{\ensuremath{\boldsymbol{#1}}}